\documentclass{article}

\usepackage{graphicx} 
\usepackage[margin=1in]{geometry} 

\usepackage{comment}
\usepackage{bbm}
\usepackage[dvipsnames, table]{xcolor}
\usepackage{mathtools} 
\usepackage{natbib}
\usepackage{graphicx} 
\usepackage{listings}
\usepackage{enumerate}
\usepackage{scalerel} 
\usepackage{amsmath}
\usepackage{amssymb}
\usepackage{amsfonts}
\usepackage{booktabs}
\usepackage{multirow}
\usepackage{threeparttable}
\usepackage{subcaption}

\usepackage{float}
\usepackage{caption}
\usepackage{colortbl}
\newcommand{\nothere}[1]{}

\usepackage{orcidlink}

\title{Analysis of Higher Education Dropouts Dynamics through Multilevel Functional Decomposition of Recurrent Events in Counting Processes}
\author{Alessandra Ragni$^{*}$\orcidlink{0000-0002-3647-7340}, Chiara Masci$^{*}$\orcidlink{0000-0002-9208-3194}, Anna Maria Paganoni$^{*}$\orcidlink{0000-0002-8253-3630} \\
\small{$^*$MOX, Department of Mathematics, Politecnico di Milano,} \\
\small{Piazza Leonardo da Vinci 32, 20133, Milano, Italy}}
\date{}

\begin{document}

\maketitle

\abstract{
This paper analyzes the dynamics of higher education dropouts through an innovative approach that integrates recurrent events modeling and point process theory with functional data analysis. We propose a novel methodology that extends existing frameworks to accommodate hierarchical data structures, demonstrating its potential through a simulation study. Using administrative data from student careers at Politecnico di Milano, we explore dropout patterns during the first year across different bachelor's degree programs and schools.
Specifically, we employ Cox-based recurrent event models, treating dropouts as repeated occurrences within both programs and schools. Additionally, we apply functional modeling of recurrent events and multilevel principal component analysis to disentangle latent effects associated with degree programs and schools, identifying critical periods of dropout risk and providing valuable insights for institutions seeking to implement strategies aimed at reducing dropout rates.}


\vspace{0.3cm}


\section{Introduction}

The higher education system is worldwide affected by high dropout rates. 
In this context, \lq\lq dropout" refers to students leaving the university world without completing their degree. From the perspective of a single university, dropout occurs when a student exits their academic program before earning the final qualification \citep{tinto1982defining}.
Despite efforts by European governments to expand access to higher education, ensuring successful degree completion remains a challenge, and dropout rates persist at around 30\% across OECD member countries \citep{oecd2019education}.
%
In Italy, this issue is particularly pronounced, with a significant proportion of students discontinuing their studies, often within the first two years of enrollment. More than half of those who begin higher education fail to complete their degrees \citep{aina2018economics}.
Indeed, the percentage of adults with a higher education degree in Italy is below the OECD average \citep{oecd2019education, cannistra2024}.
These high dropout rates not only lower the average skill levels of the workforce \citep{atzeni2022drop}, but they are also linked to a growing wage-skill gap \citep{katz1992changes}. 

From an institutional perspective, high dropout rates represent a waste of resources. 
In fact, the long-term returns — both in terms of human capital development and the credentials awarded — are lost when students exit without completing their degrees, despite the considerable investments made by universities in teaching, recruitment, and student support.
As a result, reducing and analyzing university dropout rates has become a critical challenge for higher education institutions. 

What makes managing this issue even more complex is the significant variation in dropout behavior across degree programs and schools. Even within the same university, dropout patterns differ widely across academic disciplines. For instance, some programs may experience higher dropout rates during early semesters due to difficult coursework, while others might see students leave later in their studies, near graduation. Additionally, dropout rates can vary between schools within the same univesity, influenced by factors such as faculty engagement, availability of student support services, and workload. 

In this paper, we analyze administrative data from Politecnico di Milano (PoliMi) to examine dropout patterns across its bachelor's degree programs. PoliMi comprises four distinct schools: Architecture, Design and Engineering, further divided into the School of Civil, Environmental, and Land Management Engineering and the School of Industrial and Information Engineering, offering 23 different undergraduate programs, referred to as degree courses or simply courses. Our focus is on understanding dropout rates across these programs, exploring how they vary by both degree program and school over the first-year span.

Our approach builds on methodologies that integrate recurrent events modelling, point process theory, and functional data analysis, extending the techniques proposed by \cite{baraldo2013outcome} and \cite{spreafico2021functional}. In these studies, hospital readmissions and drug consumption over time are analyzed to predict outcomes related to heart failure telemonitoring in the former and time-to-death in the latter. 
We extend and generalize this framework to account for the hierarchical structure of the data \citep{pinheiro2000linear}, enabling a more detailed exploration of dropout dynamics across different academic units. 
Specifically, the analysis comprises two phases. 
In the first phase, we utilize historical dropout data to fit a counting process model \citep{daley2002introduction}, enabling us to compute the realized trajectories of the cumulative hazard process (\textit{compensators)} underlying the dropout counting process.
While many alternatives are available for the modelization of recurrent events \citep{amorim2015modelling}, such as extensions of Cox models (see, for instance, the Prentice, Williams and Peterson model \citep{prentice1981regression}, Wei, Lin and Weissfeld model \citep{wei1989regression}, frailty models \citep{therneau2000cox}), models for the mean number of events or their occurrence rate \citep{lin2000semiparametric, diao2014statistical}, multi-state models \citep{andersen2002multi}, and virtual (effective) age models \citep{kijima1988periodical, pena2004models, beutner2021review},
we employ the Andersen-Gill (AG) model \citep{andersen1982cox}. This choice follows \cite{spreafico2021functional}, the most recent research in this context. 
The AG model extends the Cox proportional hazards model by incorporating the increments in event counts over time, assuming that correlations between event times can be explained by prior occurrences, as well as through the specification of appropriate time-varying covariates, such as the count of previous occurrences \citep{amorim2015modelling}. This allows us to represent these events as non-stationary stochastic counting processes that may depend on specific characteristics or labels, referred to as \textit{marks} \citep{daley2002introduction, spreafico2021functional}.
At this stage, the longitudinal trajectory of instantaneous dropout risk over time within a degree program is treated as a function, and functional data analysis techniques \citep{ramsay2005principal} are employed to extract insights from repeated dropout events as two-level functional covariates. 
These covariates are derived through dimensionality reduction using Multilevel Functional Principal Component Analysis (MFPCA) \citep{di2009multilevel, cui2023fast}, preserving most of the historical information while effectively managing variability across two levels.
In this first phase, our aims are twofold:
(i) to reconstruct the dropout curve by modeling dropout intensity as a counting process, capturing the temporal dynamics of dropout risk in terms of cumulative hazard for each degree program, and (ii) to decompose this dropout curve into contributions from both the degree program and school levels using MFPCA, highlighting how different academic units influence the evolution of dropout risk.
In the second phase, we adopt a predictive framework to investigate how these covariates influence the subsequent risk of dropout among students, incorporating information specific to the dropout risk associated with each faculty or school.
The aim of this second phase is to assess the predictive value of the extracted functional covariates in forecasting future dropout events at the student level. 

Previous studies focused on PoliMi data have addressed various aspects of student dropout prediction and quantification \citep{cannistra2022early, romani2023, diaz2024predicting, masci2024modelling}, with some of them specifically examining the impact of grouping factors — such as degree programs — on the time to dropout within the first few semesters of enrollment up to the full three years of bachelor degree.
The tools commonly employed in this context include shared frailty Cox proportional hazard models \citep{cook2007statistical, kleinbaum1996survival}, where the frailty term represents a constant factor shared among clusters (e.g., degree programs), which affects the baseline hazard multiplicatively, accounting for unobserved heterogeneity within clusters and allowing for a more detailed understanding of the dropout risk across different academic programs.
Our study extends previous research by offering a more refined analysis of dropout behaviour over time, specifically examining how dropout dynamics evolve both within and between degree programs and schools. A key advancement is the incorporation of the dropout history into the predictive framework which, unlike the shared frailty model that simplifies this information into a single measure, allows for a more detailed and interpretable analysis of dropout patterns, offering institutions tools for developing targeted strategies aimed at reducing dropout rates.
In this perspective, two key elements of novelty need to be highlighted.
First, we introduce the use of FMPCA to decompose the dropout curve constructed from the compensator function. This novel application allows us to separate the contributions of different academic units (e.g., programs and schools) to the overall dropout dynamics, providing a richer understanding of how these hierarchical factors influence dropout risk over time.
Second, and more importantly, our approach represents a completely new perspective in the dropout literature, where most studies focus on classification-based predictive models or, in more sophisticated cases, time-to-event models, almost exclusively at the student level (see, for instance, \cite{arulampalam2004hazard, plank2008high, min2011nonparametric, gury2011dropping, vallejos2017bayesian, patacsil2020survival}, and \cite{masci2024modelling} for a discussion). In contrast, our approach models dropout dynamics at the degree course level, capturing historical temporal patterns that can later be included for predictions at student level.

The paper is structured as follows. In Section \ref{sec:dataset} we introduce the PoliMi dataset, detailing the cohort selection and study design. 
Section \ref{sec:methodology} outlines the employed methodology, providing a recap of the framework and extending it within a multilevel context. 
Section \ref{sec:sim_study} reports on a simulation study that illustrates the application of the proposed methodology within the multilevel framework.
Section \ref{sec:case_study} presents the results obtained from applying the proposed methodology to the PoliMi case study. Discussion and concluding remarks are provided in Section \ref{sec:discussion}.

\section{Dataset}
\label{sec:dataset}
The data employed in this study were obtained from the administrative records of PoliMi, which collect the academic progress of students enrolled in bachelor's degree programs \citep{mussida2022computational}. These records encompass various aspects of students' academic \textit{careers}, including enrollment and end-of-study dates, any changes in their enrolled degree programs, and eventually incidents of dropout. Additionally, information regarding the student's history of passed and attempted \textit{exams} at different time points (semesters) is contained, including credits earned within the European Credit Transfer and Accumulation System (ECTS) and weighted Grade Point Average (GPA).
In this section, we delineate the cohort selection criteria for our study (Subsections \ref{sec:dataset_cohortselection}) alongside the study design (Subsection \ref{sec:dataset_studydesign}).

\subsection{Cohort selection}
\label{sec:dataset_cohortselection}
For our analysis, we focus on bachelor's students enrolled in academic years 2016/2017 and 2017/2018, who maintained a consistent degree program throughout their academic paths.
We assume that students enrolled in the 2017/2018 academic year were only marginally impacted by Covid-19, which occurred during the last semester of their final year.

We exclude students who graduated in under 1000 days (the minimum duration for a bachelor's degree at PoliMi) and omit fully remote or single-cycle degree programs, as the analysis focuses solely on traditional bachelor's degrees.


In the first phase of the analysis, we utilize data from students with \texttt{career\_start\_ay} = \lq 2016' to construct the compensators. For these students, we track the dropout events occurring within each \texttt{course} and \texttt{school} during the first three semesters since enrollment. In the second phase, we shift to the 2017 cohort, using data from the end of the first semester and historical information to enhance dropout risk predictions.
Table \ref{tab:variables_comp} provides an overview of the key variables used in the second phase, organized into four categories:
\begin{itemize}
    \item Variables measured at enrollment capture essential demographic and background characteristics. These include geographic origin (\texttt{origins}, i.e., whether a student lives onsite, offsite, or commutes to Milan), gender (\texttt{gender}), and age at enrollment (\texttt{age19}, which identifies students older than 19). Socio-economic status is approximated by the university fee bracket (\texttt{income}), which classifies students based on their family's financial situation into categories such as low, medium, high, or those receiving grants. Educational background is represented by the type of high school attended (\texttt{highschool\_type}). Lastly, the PoliMi admission test score (\texttt{admission\_score}) reflects academic readiness at the time of university entry, although students may take this test up to a year prior to their actual enrollment.
    \item Variables measured at the end of the first semester focus on academic progress, particularly the number of credits earned (\texttt{ECTS1sem}), a key predictor of dropout risk.
    \item Grouping factors include the undergraduate program (\texttt{course}) and broader organizational structure (\texttt{school}).
    \item The outcome variable (\texttt{dropout3y}) indicates whether a student dropped out within three years, after the first semester.
\end{itemize}


\subsection{Study design}
\label{sec:dataset_studydesign}

We label as \textit{dropouts} the students who dropped out between the end of the first and sixth semester (\textit{follow-up}), while as \textit{censored} all the other students that dropped out after three years from the enrollment, who graduated or who had an active career at the end of the third year.

With focus on students enrolled in \texttt{career\_start\_ay} = \lq 2016', we include a three-semesters \textit{observation period}, denoted by $S = [T_0, T_1]$ with $ T_0 = \text{\lq} \texttt{career\_start\_ay}/10/01 \text{\rq}$ 
and $ T_1 = \text{\lq} \texttt{career\_start\_ay} + 2/03/01 \text{\rq}$. 
The date of October 1st is chosen to exclude students who dropped out within the first two weeks, potentially due to waiting for other university entrance test results\footnote{At PoliMi, lectures typically begin in mid-September.}, to ensure they do not affect the analysis.
During this period, dropouts of students enrolled in the chosen \texttt{career\_start\_ay} are monitored and analyzed based on various grouping factors (\texttt{course} and \texttt{school}).

Following this, the focus is moved to the cohort of students with \texttt{career\_start\_ay} = \lq \texttt{2017}', particularly at the end of the first semester, and the primary outcome of interest is the binary variable \texttt{dropout3y} indicating whether a student dropped out within three years from enrollment.
To predict this outcome, we use data collected at enrollment, at the end of the first semester, and at the level of grouping factors. The choice is based on previous studies' results \citep{masci2024modelling}, which indicate that the optimal prediction window occurs within the first few semesters, as the inclusion of data from later semesters provides minimal improvement in accuracy. Moreover, incorporating hierarchical information has been shown to enhance predictive performance, with the number of credits earned by the end of the first semester serving as a particularly strong predictor.
Furthermore, the model incorporates information derived from the analysis of the previous academic year's data, adding valuable historical context to enhance predictive accuracy.

\begin{table}
\resizebox{\textwidth}{!}{%
\begin{tabular}{@{}lll@{}}
\toprule
\textbf{Variable} & \textbf{Description}  & \textbf{Type}  \\
\midrule
\textit{Measured at enrollment} & & \\
\cmidrule{1-1}
- \texttt{studentID} & Student's unique identifier (anonymized)  & Categorical \{\texttt{1,2,\dots}\}  \\
- \texttt{origins} & Student's geographic origins  & Categorical \{\texttt{OnSite, Commuter, Offsite}\} \\
- \texttt{gender} & Student's gender  & Categorical \{\texttt{Male, Female}\} \\
- \texttt{highschool\_type} & Type of attended high school   & Categorical \{\texttt{Scientific,} \\
& &  \texttt{Classical, Others, Technical}\} \\
- \texttt{income} & University fee brakcet   & Categorical \\
& & \{\texttt{Medium, Grant, High, Low}\} \\
- \texttt{age19} & Equals 1 if student’s age at enrollment $>$ 19  & Categorical \{\texttt{0,1}\} \\
- \texttt{admission\_score} & PoliMi entrance test’s admission score  & Real number [60, 100] \\
- \texttt{career\_start\_ay} & Student's enrollment year & Categorical \{\texttt{2016, 2017}\} \\
\midrule
\textit{Measured at end of 1st semester} & & \\
\cmidrule{1-1}
- \texttt{ECTS1sem} & ECTS gained by end of 1st semester & Natural number $>0$ \\
\midrule
\textit{Grouping factors} & & \\
\cmidrule{1-1}
    - \texttt{course} & Undergraduate program & Categorical \{\texttt{P01, P02,\dots, P23}\} \\
- \texttt{school} & A larger organizational unit grouping courses &  Categorical  \{\texttt{sA, sB, sC, sD}\} \\
\midrule
\textit{Outcome} & & \\
\cmidrule{1-1}
- \texttt{dropout3y} & Equals 1 if after 1st semester & Categorical \{\texttt{0, 1}\} \\
&  a student drops within 3 years, 0 otherwise &  \\
\bottomrule
\end{tabular}
}
\begin{tablenotes}
\footnotesize
\item Note: in categorical variables, the first reported class represents the reference level.
\end{tablenotes}
\caption{Overview of the variables considered in the analysis.}
\label{tab:variables_comp}
\end{table}

\section{Methodology}
\label{sec:methodology}

In this section, we present the methodology in three consequent steps: recap on model formulation for recurrent events and compensators reconstruction (Subsection \ref{sec:modelrecevents}), compensators decomposition through multilevel principal component analysis (Section \ref{sec:comprecostrucdecomp}) and the development of a predictive model for the dropout status within three years including retrieved information (Section \ref{sec:functionallogisticmodel}).
The core methodological contribution of this work regards the extension to the multilevel setting of the decomposition of recurrent events in counting processes. 

\subsection{Recap on the model formulation for recurrent events and compensators reconstruction}
\label{sec:modelrecevents}


Let $N_{ij}(t)$, with $t \in [0, T]$, denote the stochastic process counting the dropout events observed up to time $t$, where $j=1,...,J_i$ indexes the \texttt{course}-level (lower-level) units and $i=1,...,I$ indexes the \texttt{school}-level (higher-level) units, or clusters, with the total number of lower-level units given by $\sum_{i=1}^I J_i = n$ \citep{cook2007statistical}. 
The process $N_{ij}(t)$ is adapted to the filtration $\{\mathcal{F}_{t,ij}\}_{t \in [0,T]}$, that is the history of realizations of the process itself.
Assuming $N_{ij}(t)$ is a class D submartingale, the Doob-Meyer (D-M) decomposition theorem \citep{meyer1962decomposition} states that $M_{ij}(t) = N_{ij}(t) - \Lambda_{ij}(t)$ is a zero-mean, uniformly integrable martingale. Here, $\Lambda_{ij}(t) = \int_0^t \lambda_{ij}(s) ds$ is the unique predictable non-decreasing cadlag\footnote{i.e., right-continuous with left limits.} and integrable \textit{compensator} (or cumulative hazard process), with $\lambda_{ij}(t)$ being the intensity process (or hazard function). 

Building on the formulation for marked point processes described in \cite{spreafico2021functional} and extending it to a multilevel context, the events, whose cumulative number up to a given time $t$ are recorded by the counting process $N_{ij}(t)$, can be further associated to 
additional random variables (marks) $\boldsymbol{\omega}_{ij}$ that provide further details about these events, such as the size or magnitude related to the jumps in the counting process \citep{last1995marked, daley2003introduction}.
In this framework, the conditional intensity function also depends on the mark 
$\boldsymbol{\omega}_{ij}$. Assuming conditional independence of jump times and marks, the following relationship holds
$ \lambda_{ij}(t, \boldsymbol{\omega}_{ij}) = \lambda_{ij,g}(t) \cdot f_{ij}(\boldsymbol{\omega}_{ij})$, where $\lambda_{ij,g}(t)$ is the ground intensity process of the counting process and $f_{ij}$ is the multivariate density of the marks $\boldsymbol{\omega}_{ij}$.
Proper modeling of compensators and particularly of $\lambda_{ij}(t, \boldsymbol{\omega}_{ij})$, allows for an accurate reconstruction of $N_{ij}(t)$, as $M_{ij}(t)$ represents the residual of the process in the D-M decomposition.

Several models for $\lambda_{ij}(t)$ are available in the literature on counting processes \citep{aalen2008survival, andersen2012statistical, pena2004models}. Employing the model introduced by 
\citep{andersen1982cox}, under the assumption that $f_{ij}(\boldsymbol{\omega}_{ij})$ depends on $\mathbf{z}_{ij}(t)$ (that are some time-dependent features related to the marks $\boldsymbol{\omega}_{ij}$), we get
\begin{align}
\label{Eq:AGmodels}
    \lambda_{ij}(t, \boldsymbol{\omega}_{ij}) & = Y_{ij}(t) \; \lambda_0(t) \;\text{exp}\{\boldsymbol{\beta}^T \mathbf{x}_{ij}(t)\} \;\text{exp}\{\boldsymbol{\theta}^T \mathbf{z}_{ij}(t)\} \nonumber \\ 
    & = Y_{ij}(t) \; \lambda_0(t) \;\text{exp}\{\boldsymbol{\beta}^T \mathbf{x}_{ij}(t) + \boldsymbol{\theta}^T \mathbf{z}_{ij}(t)\}
\end{align}
where $\mathbf{x}_{ij}(t)$ are the (time-dependent) column vectors of covariates of the $j^{\text{th}}$ unit in $i^{\text{th}}$ cluster, $\lambda_0(t)$ is the baseline hazard function, $Y_{ij}$ takes the role of the censoring variable (i.e. assumes value 1 when unit $j$ in cluster $i$ is under observation), $\boldsymbol{\beta}$ and $\boldsymbol{\theta}$ are $Q$- and $P$-dimensional column vectors of coefficient  and $^T$ stands for the transpose.
In particular, following \cite{spreafico2021functional}, the mark density $f_{ij}(\boldsymbol{\omega}_{ij})$ is 
incorporated into the model through the exponential term involving $\mathbf{z}_{ij}(t)$, which parametrizes the influence of the marks on the process.
$\boldsymbol{\beta}$ and $\boldsymbol{\theta}$ are estimated in the model fitting by partial likelihood maximization \citep{andersen1982cox}, while the baseline cumulative hazard $\Lambda_0(t) = \int_0^t \lambda_0(s) ds$ can be estimated through Breslow estimator \citep{breslow1975analysis} as a step-function $\hat{\Lambda}_0(t)$ and then smoothed into $\tilde{\Lambda}_0(t)$ as described in \cite{baraldo2013outcome}. 

Let now $[t_k^{(ij)}, t_{k+1}^{(ij)}]$ for $k=0,..., N_{ij}(T)$
be the intervals whose extremes are the jump times for each unit $j$ in cluster $i$, being $t_0^{(ij)} = 0$ and $t_{N_{ij}(T) + 1}^{(ij)} = T$. 
Then $\Lambda_{ij}(t) = \int_0^t \lambda_{ij}(s, \boldsymbol{\omega}_{ij}) ds $ can be estimated by approximation as follows (see computation in Appendix \ref{app:compreconstr}): 
\begin{equation}
    \hat{\Lambda}_{ij}(t) =  \sum_{k=0}^{N_{ij}(t^-)} \; \exp(\hat{\boldsymbol{\beta}}^T \mathbf{x}_{ij}(t_k^{(ij)}) + \hat{\boldsymbol{\theta}}^T \mathbf{z}_{ij}(t_k^{(ij)})) 
    \big[ \tilde{\Lambda}_0(t_{k+1}^{(ij)} \wedge t) - \tilde{\Lambda}_0(t_k^{(ij)})\big] . \label{eq:reconstruction}
\end{equation}
where $a\wedge b = \min\{a,b\}$, $N_{ij}(t^-)$ represents the number of occurrences that have happened strictly before time $t$, and $\hat{\boldsymbol{\beta}}$ and $\hat{\boldsymbol{\theta}}$ are the estimated vectors of coefficients.

\subsection{Multilevel functional principal component analysis for compensators decomposition}
\label{sec:comprecostrucdecomp}

After reconstructing compensators through a marked point process formulation for recurrent events, $\hat{\Lambda}_{ij}(t)$ can be regarded as functional data objects, allowing the application of functional data analysis techniques \citep{ramsay2005principal}.

Given the high-dimensional nature of these data and the hierarchical setting, we aim to decompose functional variability and reduce dimensionality, while getting insights. To achieve this, we apply MFPCA \citep{di2009multilevel,cui2023fast}. MFPCA integrates classical FPCA \citep{ramsay2005principal}, which selects only the relevant components of an appropriate orthonormal basis expansion, 
with standard multilevel mixed models. This approach effectively decomposes the observed data according to two levels of functional variation.
Specifically, 
from the one-way functional ANOVA \citep{di2009multilevel} follows that
\begin{align}
    \hat{\Lambda}_{ij}(t) & =  \mu(t) + Z_i(t) + W_{ij}(t) + \epsilon_{ij}(t)  \label{eq:oneway}  \\
    & = \mu(t) + \sum_{k=1}^{\infty} \xi_{ik} \phi_k^{(1)}(t) + \sum_{l=1}^{\infty} \zeta_{ijl} \phi_l^{(2)}(t)+ \epsilon_{ij}(t) \label{eq:decompos}
\end{align}
where, in Eq. (\ref{eq:oneway}), $\mu(t)$ is a fixed functional effect, $Z_i(t)$ and $W_{ij}(t)$ are mean 0 stochastic processes (uncorrelated between each other) and $\epsilon_{ij}$ is observed only when functional data are observed with errors.
Eq. (\ref{eq:decompos}) follows from Karhunen-Loève (KL) expansion \citep{karhunen1947under, loeve1948functions}, where $\phi_k^{(1)}(t)$ and $\phi_l^{(2)}(t)$ are respectively level 1 (i.e., cluster level) and level 2 (i.e., unit level) eigenfunctions (fixed functional effects), and $\xi_{ik}$ and $\zeta_{ijl}$ are respectively level 1 and 2 principal component scores (zero mean random variables, uncorrelated between each other).
Moreover, one may truncate the decomposition by pre-specifying at both levels the Percentage of Variance Explained (PVE) as explained in \cite{di2009multilevel}, resulting into
\begin{equation}
    \hat{\Lambda}_{ij}(t) \simeq \mu(t) + \sum_{k=1}^{K} \xi_{ik} \phi_k^{(1)}(t) + \sum_{l=1}^{L} \zeta_{ijl} \phi_l^{(2)}(t) + \epsilon_{ij}(t)   
    \label{eq:compensatormultileveldecomp}
\end{equation}
being $K$ and $L$ the number of principal components finally identified respectively at level 1 (cluster) and level 2 (unit). Other interesting indicators defined in \cite{di2009multilevel} are the total explained variance between-clusters and within-clusters, and the proportion of variability explained by level 1.

\subsection{Logistic regression model with functional compensators}
\label{sec:functionallogisticmodel}
The compensators decomposition in previous sections allows to extract dropout information focusing on the \textit{observation period} $S = [0,T]$, where the units are at \texttt{course}-level $j$, clustered within \texttt{school}-level $i$.
As a last step, we include this functional information into a logistic regression model which considers another cohort, where the units are now at \texttt{studentID}-level $h$, for $h = 1,..., H_{ij}$, nested within level $j$ (\texttt{course}-level) again nested within level $i$ (\texttt{school}-level), so that $n_{\text{tot}} = \sum_{i=1}^{I}\sum_{j=1}^{J_i} H_{ij}$ is the total number of units.
Let $Y_{ijh} \sim \text{Bernoulli}(p_{ijh})$
be the binary variable indicating the output \texttt{dropout3y}.
Then
\begin{align}
    \text{logit}(p_{ijh}) & = \boldsymbol{\gamma}^T \mathbf{w}_{ijh} + \int_S  \Lambda_{ij}(s)  \alpha(s) ds \nonumber \\
    & \simeq \boldsymbol{\gamma}^T \mathbf{w}_{ijh} + \int_S  \Big[ \sum_{k=1}^{K} \xi_{ik} \phi_k^{(1)}(s) + \sum_{l=1}^{L} \zeta_{ijl} \phi_l^{(2)}(s) \Big] \alpha(s) ds  \nonumber \\
    & = \boldsymbol{\gamma}^T \mathbf{w}_{ijh} + \sum_{k=1}^{K} \xi_{ik} \alpha_k^{(1)} + \sum_{l=1}^{L} \zeta_{ijl} \alpha_l^{(2)}
\end{align}
for $i=1,...,I$, $j=1,...,J_i$ and $h=1,...H_{ij}$ where $\boldsymbol{\gamma}$ is a $q$-dimensional vector of parameters to be estimated, $\mathbf{w}_{ijh}$ is a vector of covariates available at unit level $h$, $\alpha: S \rightarrow \mathbb{R}$ is a functional parameter and $\text{logit}(x) := \ln\big(\frac{x}{1-x}\big)$.
The second line follows from Eq. (\ref{eq:compensatormultileveldecomp})
and last equality is given by rewriting $\alpha (s)$ according to different representations into the two different orthonormal bases $\phi_k^{(1)}$ and $\phi_l^{(2)}$, thanks to the orthonormality property; the subscripts are added in order to distinguish the two projections.

\section{Simulation Study}
\label{sec:sim_study}

In this section, we aim to demonstrate the effectiveness of the methodology described above, in particular in Sections \ref{sec:modelrecevents} and \ref{sec:comprecostrucdecomp}. After simulating unit-level intensities with shapes based on specific similarities within clusters and generating the Non-Homogeneous Poisson Processes (NHPPs) from these intensities, we show that compensators reconstruction using AG models effectively recovers the within-cluster similarities. 

Specifically, in Subsection \ref{sec:sim:dgp}, we begin by simulating intensities $\lambda_{ij}(t)$ following a similar methodology to \cite{cui2023fast, di2009multilevel}. We employ a one-way functional ANOVA model to capture similarities within clusters and integrate $\Lambda_{ij}(t) = \int_{0}^t \lambda_{ij}(u) du$ to obtain compensator-like shapes, translating the intensity functions into cumulative hazard functions, and we simulate event times from the $\lambda_{ij}(t)$.
In Subsection \ref{sec:nhpp_agfitting}, we fit AG models to the simulated event data, and reconstruct the compensators $\hat{\Lambda}_{ij}(t)$.
Finally, in Subsection \ref{sec:mfpca_simulation} we evaluate the consistency of the information captured by MFPCA before and after NHPPs extraction. We aim to show that cumulative hazard reconstruction using AG models preserves the essential information captured by the MFPCA.

\subsection{Data Generating Process}
\label{sec:sim:dgp}

Let $\lambda_{ij}(t)$ be an intensity function measured over a continuous variable $t \in [0, 1]$ for observation $j$ within cluster $i$, for $j=1,\dots,J_i$ and $i=1,\dots,I$, generated by a  modified one-way functional ANOVA model \citep{morris2003wavelet} as follows:
\begin{equation}
    \lambda_{ij}(t) := \mu(t) + 2\cdot i \cdot \Big( \sum_{k=1}^{K} \xi_{ik} \, \phi_{k}^{(1)}(t) + \sum_{l=1}^{L} \zeta_{ijl} \, \phi_{l}^{(2)}(t) + \epsilon_{ij}(t) \Big)
    \label{eq:simulationintensities}
\end{equation}
where $\mu(t) = 200$, $\xi_{ik} \sim \mathcal{N}(0, \lambda_{k}^{(1)} )$, $\zeta_{ijl} \sim \mathcal{N}(0, \lambda_{l}^{(2)})$ and $\epsilon_{ij} \sim \mathcal{N}(0, \sigma^2)$. 

For our data generation, we draw inspiration from the simulation study described in Section 4 of \cite{di2009multilevel}, introducing a few modifications to make the generation of the intensities more suitable for our specific context.
Firstly, we include a constant $\mu(t)$ to increase the frequency of events in the NHPP generated by $\lambda_{ij}(t)$. Additionally, we scale level 1 and 2 components by a cluster-dependent constant, enhancing the differentiation between-groups and, consequently, the cumulative intensities. Lastly, we assign a higher standard deviation to the true eigenvalues at level 1 compared to level 2, further distinguishing clusters and reducing within-cluster variability.

The decision to simulate the intensities rather than directly simulating the cumulative hazard function stems from the specific characteristics required for the cumulative hazard (increasing monotonicity and ensuring that $\Lambda_{ij}(0) = 0$). Simulating scores from a normal distribution while maintaining these properties is not possible. Therefore, we opt to simulate the intensities to ensure these essential characteristics are preserved.

\begin{figure}
\centering
\subfloat[$\lambda_{ij}(t)$]{
\includegraphics[width = 0.45\columnwidth]{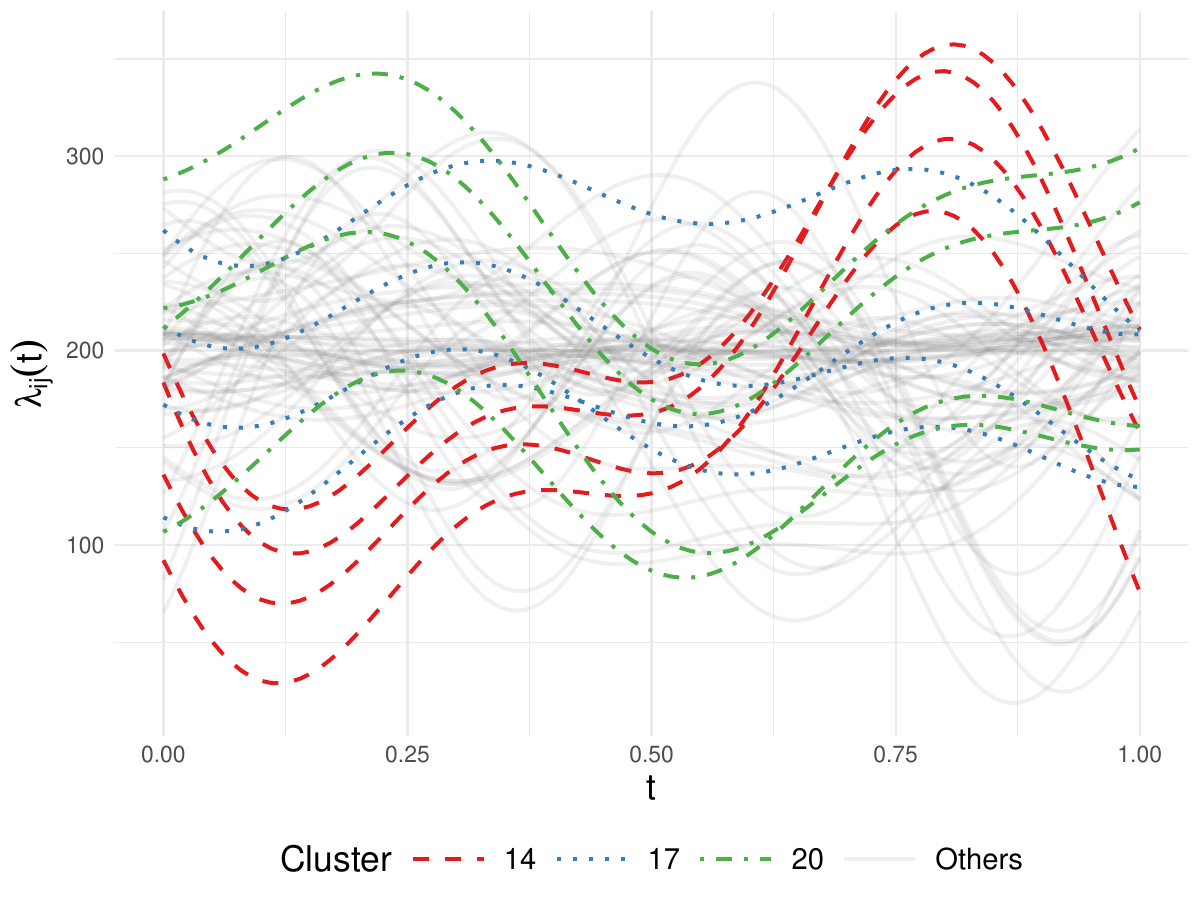}} 
\hspace{0.1cm}
\subfloat[$\Lambda_{ij}(t) = \int_{0}^t \lambda_{ij}(u) du$]{
\includegraphics[width = 0.45\columnwidth]{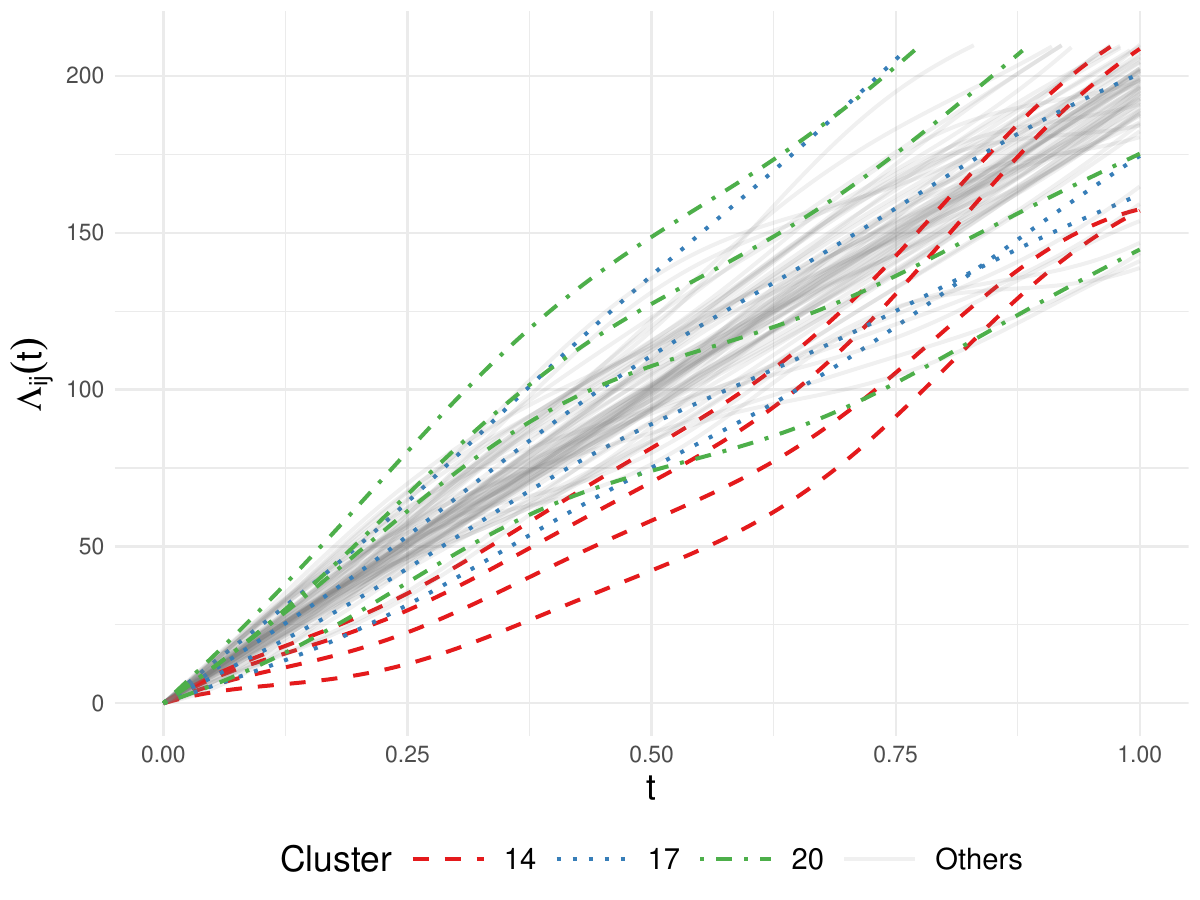} }
\caption{Simulated intensities $\lambda_{ij}(t)$ (i) and cumulative hazards $\Lambda_{ij}(t)$ (ii), with clusters 14, 17, and 18 highlighted using different colors and line types due to their outlying shapes,  which stand out from the general patterns observed in other clusters.}
\label{fig:lambdaLambda}
\end{figure}

We assume $I = 20$ clusters, $J = 4$ units and $K = L = 4$. 
The chosen value of the eigenvalues are $\lambda_k^{(1)} = 0.9^{k-1}$ for $k=1,...,K$ and $\lambda_l^{(2)} = 0.2^{l-1}$ for $l=1,...,L$, while the chosen value of the eigenfunctions, chosen following \cite{di2009multilevel, cui2023fast}, are
\begin{equation}
    \{ \phi_{1}^{(1)}(t), \phi_{2}^{(1)}(t), \phi_{3}^{(1)}(t), \phi_{4}^{(1)}(t)\} = \{ \sqrt{2} \, \text{sin}(2\pi t), \sqrt{2} \, \text{cos}(2\pi t), \sqrt{2} \, \text{sin}(4\pi t), \sqrt{2} \, \text{cos}(4\pi t) \}
    \label{eq:eigenfunctionslevel1}
\end{equation}
\begin{equation}
    \{ \phi_{1}^{(2)}(t), \phi_{2}^{(2)}(t), \phi_{3}^{(2)}(t), \phi_{4}^{(2)}(t) \} = \{ 1, \sqrt{3} \, (2t-1), \sqrt{5} \, (6 t^2 - 6t +1), \sqrt{7} \, (20t^3 -30t^2 + 12t -1) \}
    \label{eq:eigenfunctionslevel2}
\end{equation}
at levels 1 and 2, respectively.
Moreover, we assume $\mu(t) = 100$ and $\sigma=0$ (no noise). Afterwards, we compute the cumulative hazard function as $\Lambda_{ij}(t) = \int_{0}^t \lambda_{ij}(u) du$.
In Figure \ref{fig:lambdaLambda}, we illustrate the simulated intensities and cumulative hazards.

Following \cite{pasupathy2010generating}, through the thinning method we simulate event times for a NHPP over $[0, 1]$, where the process is characterized by the time-varying intensity function $\lambda_{ij}(t)$.

\subsection{Results}

\subsubsection{Compensators estimation}
\label{sec:nhpp_agfitting}

At this point, we apply the pipeline described in Section \ref{sec:modelrecevents}, where in the AG model in Eq. (\ref{Eq:AGmodels}) we employ the number of events recorded up to that time interval for unit $j$ in cluster $i$ as a time-dependent covariate.

\begin{figure}
\centering
\subfloat[$\hat{\Lambda}_{0}(t)$]{
\includegraphics[width = 0.45\columnwidth]{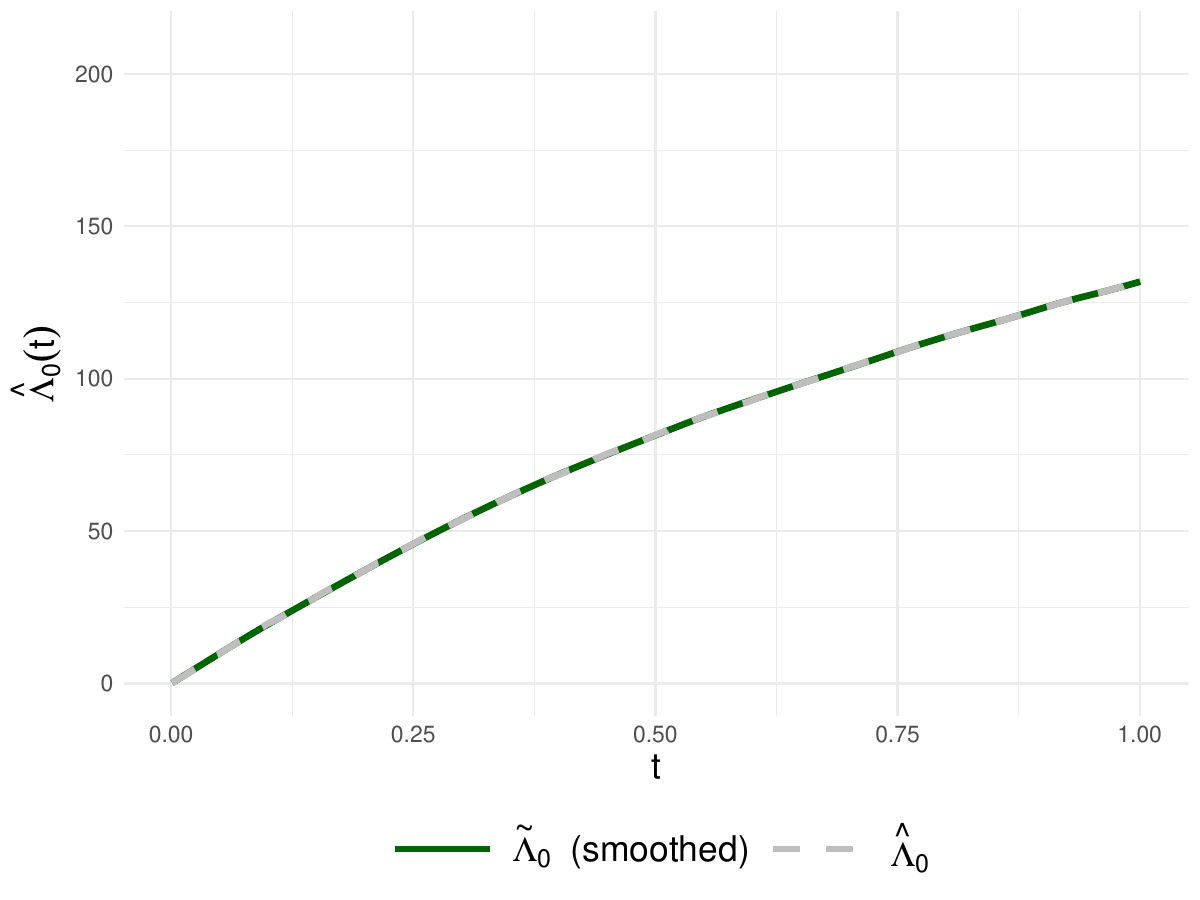}} 
\hspace{0.1cm}
\subfloat[$\hat{\Lambda}_{ij}(t)$]{
\includegraphics[width = 0.45\columnwidth]{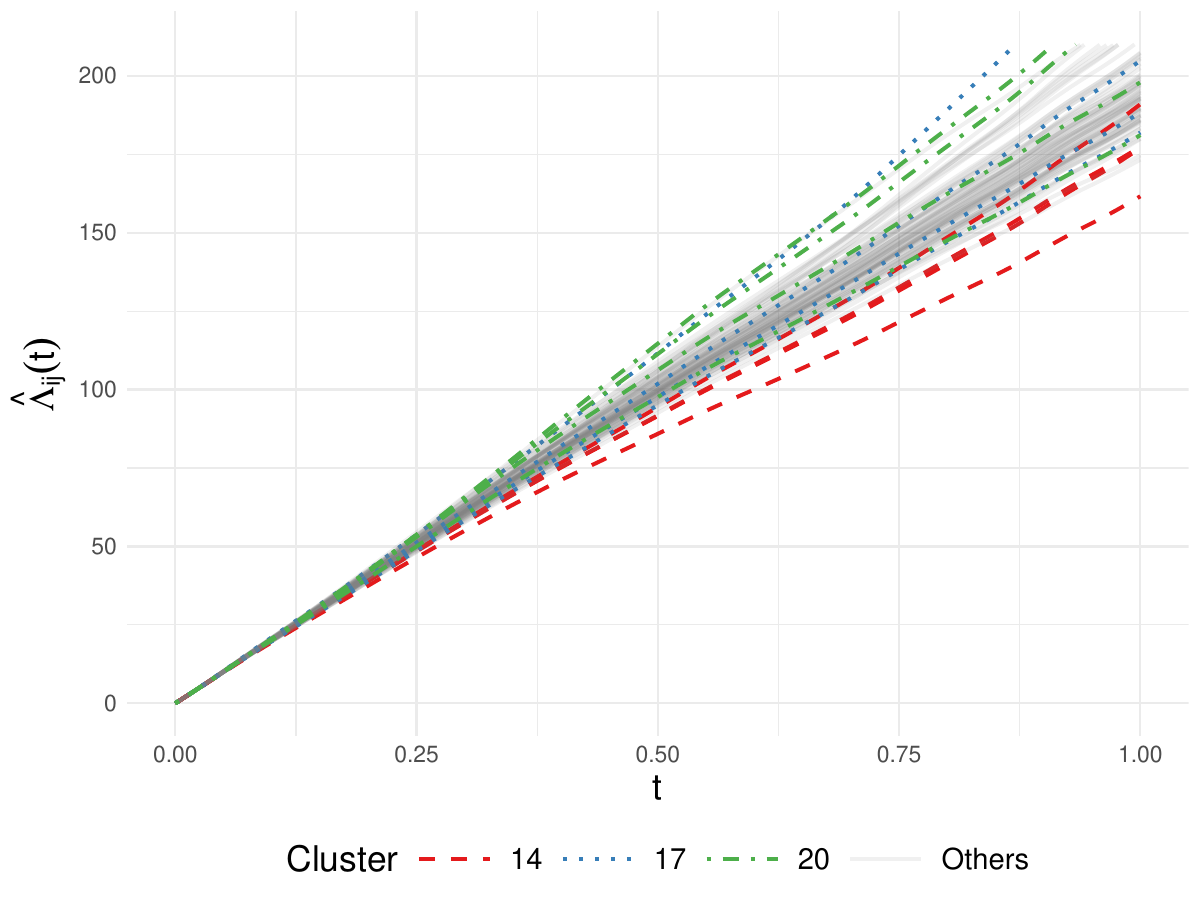} }
\caption{Estimated $\hat{\Lambda}_{0}(t)$ (i) and $\hat{\Lambda}_{ij}(t)$ (ii). In (ii), clusters 14, 17, and 18 are highlighted using different colors and line types due to their outlying shapes, which stand out from the general patterns observed in other clusters.}
\label{fig:LambdaPREPOST}
\end{figure}

The baseline cumulative hazard, $\hat{\Lambda}_0(t)$, along with $\tilde{\Lambda}_0(t)$ is estimated. Additionally, $\hat{\Lambda}_{ij}(t)$ is derived according to Eq. (\ref{eq:reconstruction}). These functions are depicted in Figure \ref{fig:LambdaPREPOST}.
By visual inspection, we observe some information loss due to the stochastic nature of NHPPs in the simulation of event times. Nonetheless, cluster behaviours remain distinguishable and can still be effectively recognised and characterized.

\subsubsection{Multilevel functional principal component analysis}
\label{sec:mfpca_simulation} 

\begin{figure}
\centering
\includegraphics[width = 0.45\columnwidth]{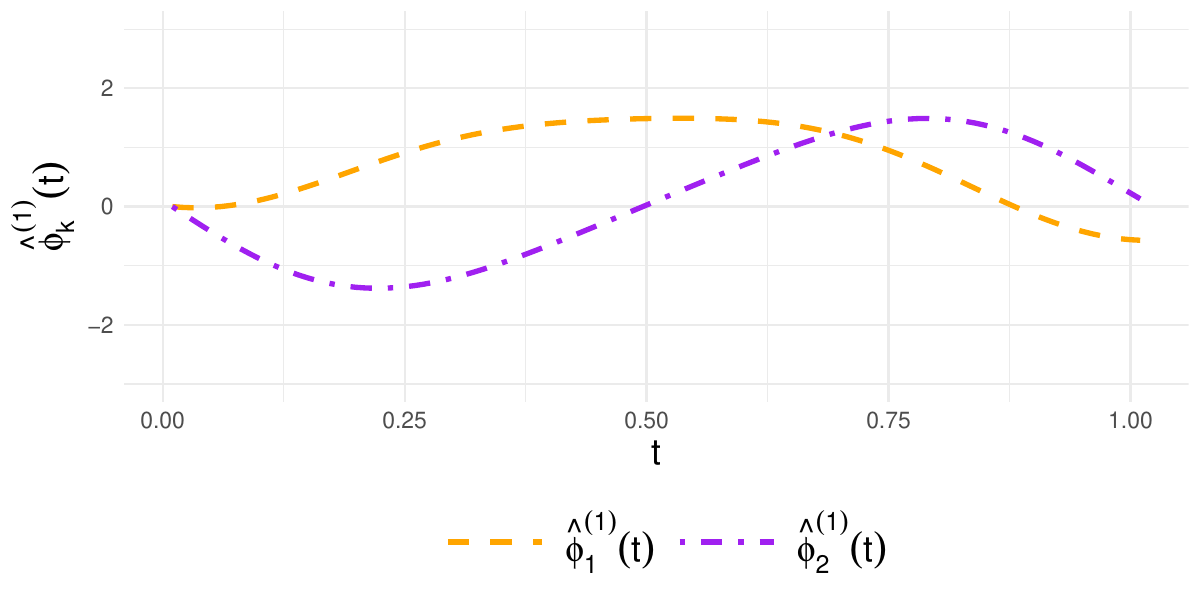} \hspace{1cm}
\includegraphics[width = 0.45\columnwidth]{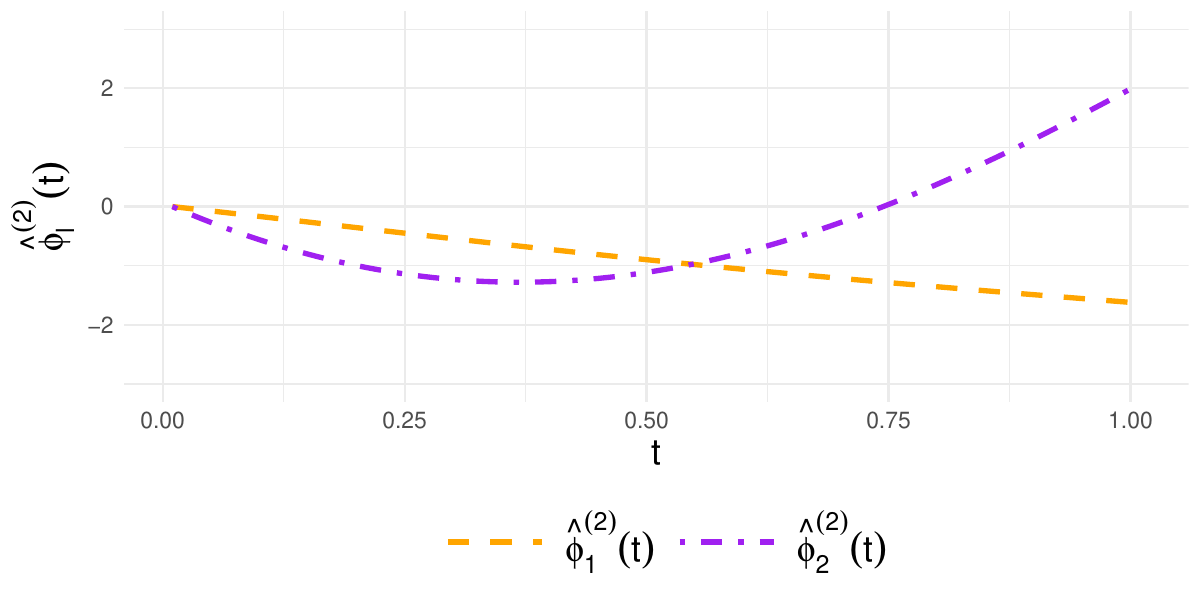} 
\caption{First two eigenfunctions at levels 1 and 2 (left and right panels, respectively), computed from simulated $\Lambda_{ij}(t)$ represented in Figure \ref{fig:lambdaLambda} (ii).}\label{fig:LambdaEigenFun}
\end{figure}

As final step of our simulation study, we implement the decomposition described in Section \ref{sec:comprecostrucdecomp}. 
We recall that in Eq. (\ref{eq:simulationintensities}) we simulate intensities $\lambda_{ij}(t)$ employing the eigenfunctions in Eq. (\ref{eq:eigenfunctionslevel1}-\ref{eq:eigenfunctionslevel2}). However, our primary interest lies in $\Lambda_{ij}(t)$.
Therefore, we first decompose $\Lambda_{ij}(t)$ using Eq. (\ref{eq:compensatormultileveldecomp}). 
At this stage, we obtain 4 functional principal components at level 1 and 2 at level 2. To determine the number of principal components at both levels, we set the PVE to $0.99$, following the default setting in the \texttt{mfpca.face} function from \cite{cui2023fast}.
In Figure \ref{fig:LambdaEigenFun}, 
we show results of MFPCA on functional compensators related to the first and second principal components, respectively for levels 1 and 2. 

\begin{figure}
\centering
\includegraphics[width = 0.45\columnwidth]{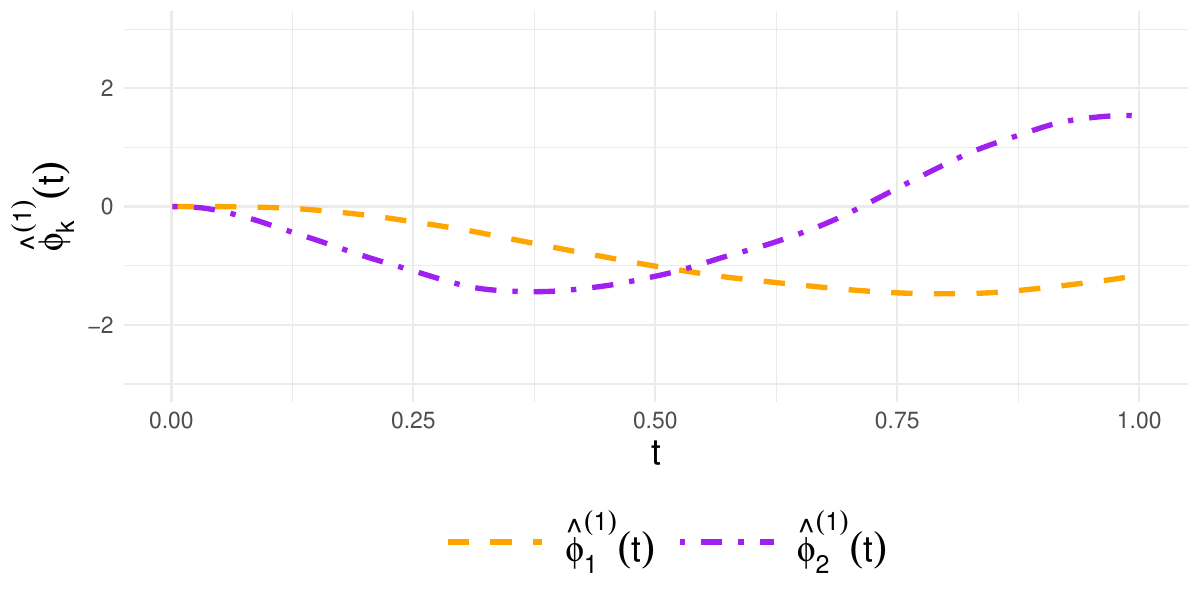} \hspace{1cm}
\includegraphics[width = 0.45\columnwidth]{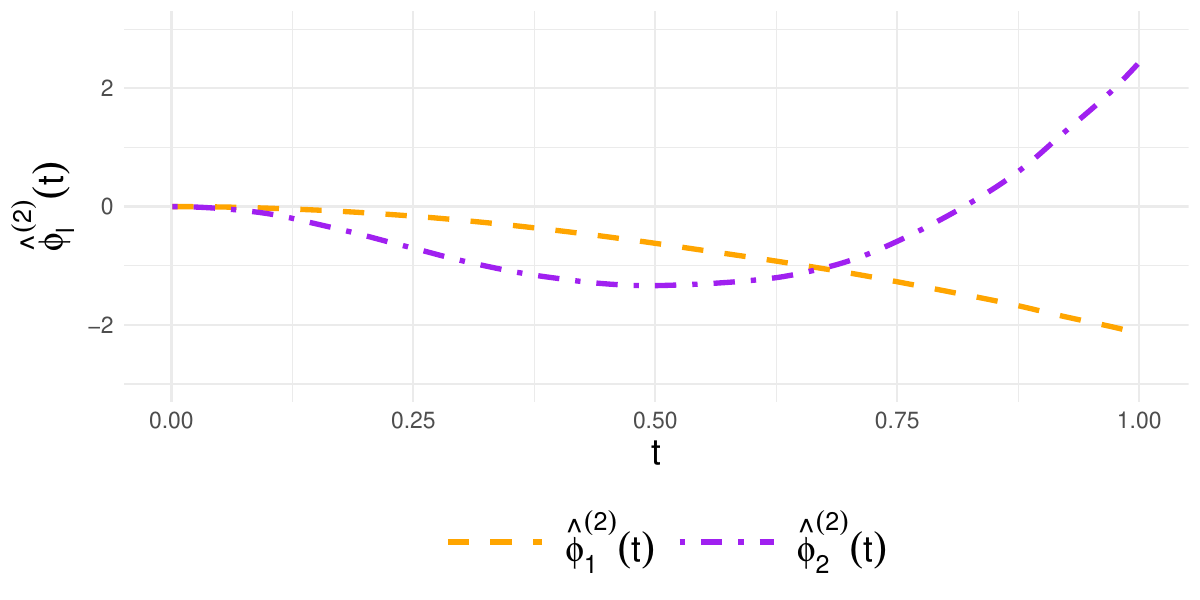} 
\caption{First two eigenfunctions at levels 1 and 2 (left and right panels, respectively), computed from reconstructed $\hat{\Lambda}_{ij}(t)$ represented in Figure \ref{fig:LambdaPREPOST} (ii).}
\label{fig:LambdaEigenFunPOST}
\end{figure}

On the other hand, after the simulation of the NHPP as described in previous section and having computed $\hat{\Lambda}_{ij}(t)$ according to Eq. (\ref{eq:reconstruction}), we apply the same multilevel functional decomposition to $\hat{\Lambda}_{ij}(t)$. Here, we obtain 3 functional principal components at level 1 and 2 at level 2 and of degree course, the magnitude of the eigenvalues reduce.
However, if we analyse the eigenfunctions reported in Figure \ref{fig:LambdaEigenFunPOST}, 
similar pattern can be observed, both for levels 1 and 2. 

In general, we observe that the reconstructed compensators closely match the original simulated shapes, despite some information loss due to the sampling and fitting processes. Results indicate that cumulative hazard reconstruction using AG models preserves the essential information captured by MFPCA and could be further enriched by incorporating additional application-specific covariates. This demonstrates that for a NHPP with cluster-similar intensities, the cumulative hazard reconstruction using AG models effectively retrieves the simulated shapes, retaining the crucial information captured by MFPCA prior to process extraction.

\section{Case Study}
\label{sec:case_study}

We apply the proposed methodology to a case study involving the administrative dataset of PoliMi. First, we present the compensators reconstruction and decomposition at degree \texttt{course} and  \texttt{school} levels (Subsection \ref{sec:compensators}), then effectively implement a predictive model at \texttt{studentID}-level (Subsection \ref{sec:prediction}).


\subsection{Compensators reconstruction and decomposition}
\label{sec:compensators}

For the analysis of dropouts as a marked point process, after cohort selection described in Subsection \ref{sec:dataset_cohortselection} and having filtered the data to focus on a specified academic year (in our case, \texttt{career\_start\_ay} = \lq 2016'), 
we establish start and stop dates for each dropout event that happened on distinct days and enumerate the cumulative occurred dropout distinct days (\texttt{enum}), as well as the number of events (\texttt{dropout\_count}) standardized by the number of students enrolled in that course, that will perform as the mark of the counting process.
Afterwards, these covariates are employed for the AG model for recurrent events describing the dropouts. 

\begin{figure}
\centering
\subfloat[Baseline cumulative hazard $\hat{\Lambda}_{0}(t)$]{\includegraphics[width = 0.48\columnwidth]{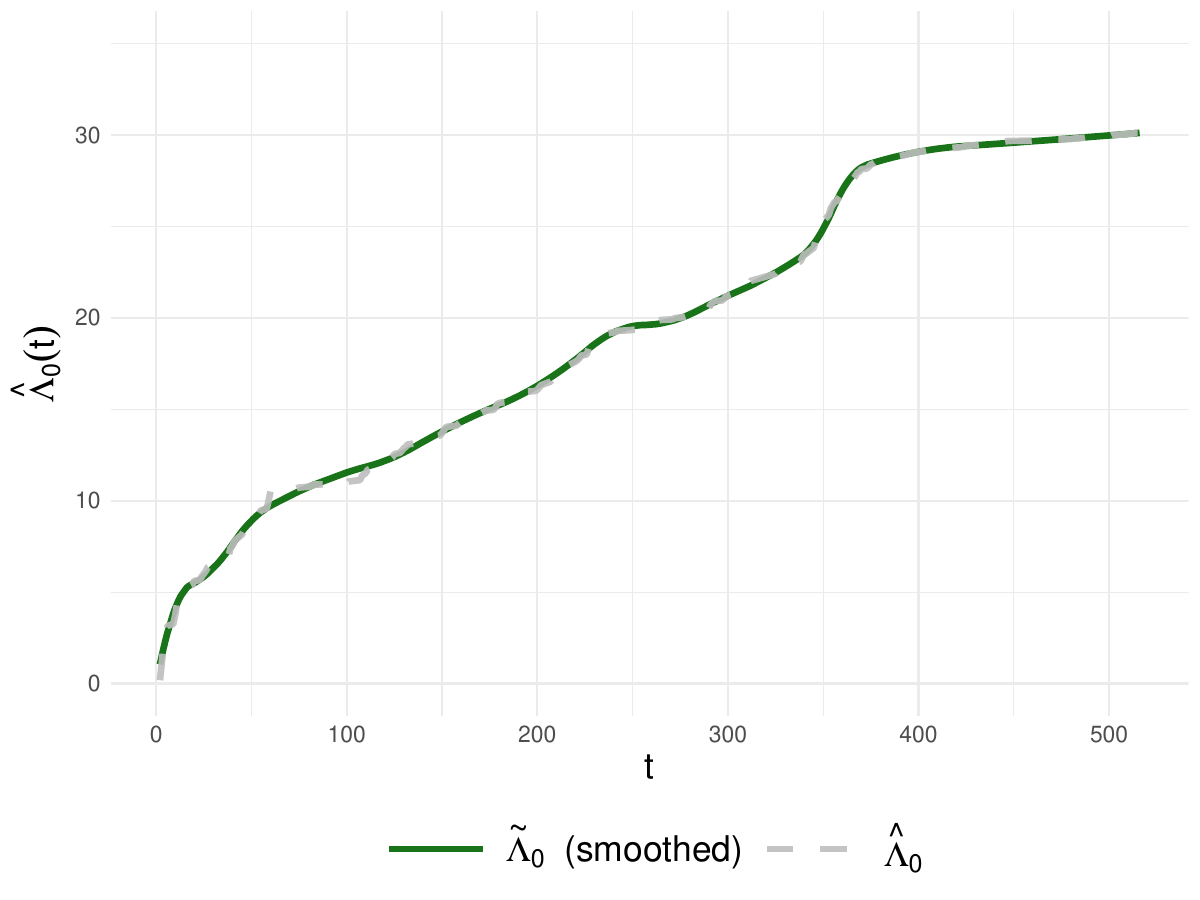}} \hspace{0.1cm}
\subfloat[Reconstructed compensators $\hat{\Lambda}_{ij}(t)$]{\includegraphics[width = 0.48\columnwidth]{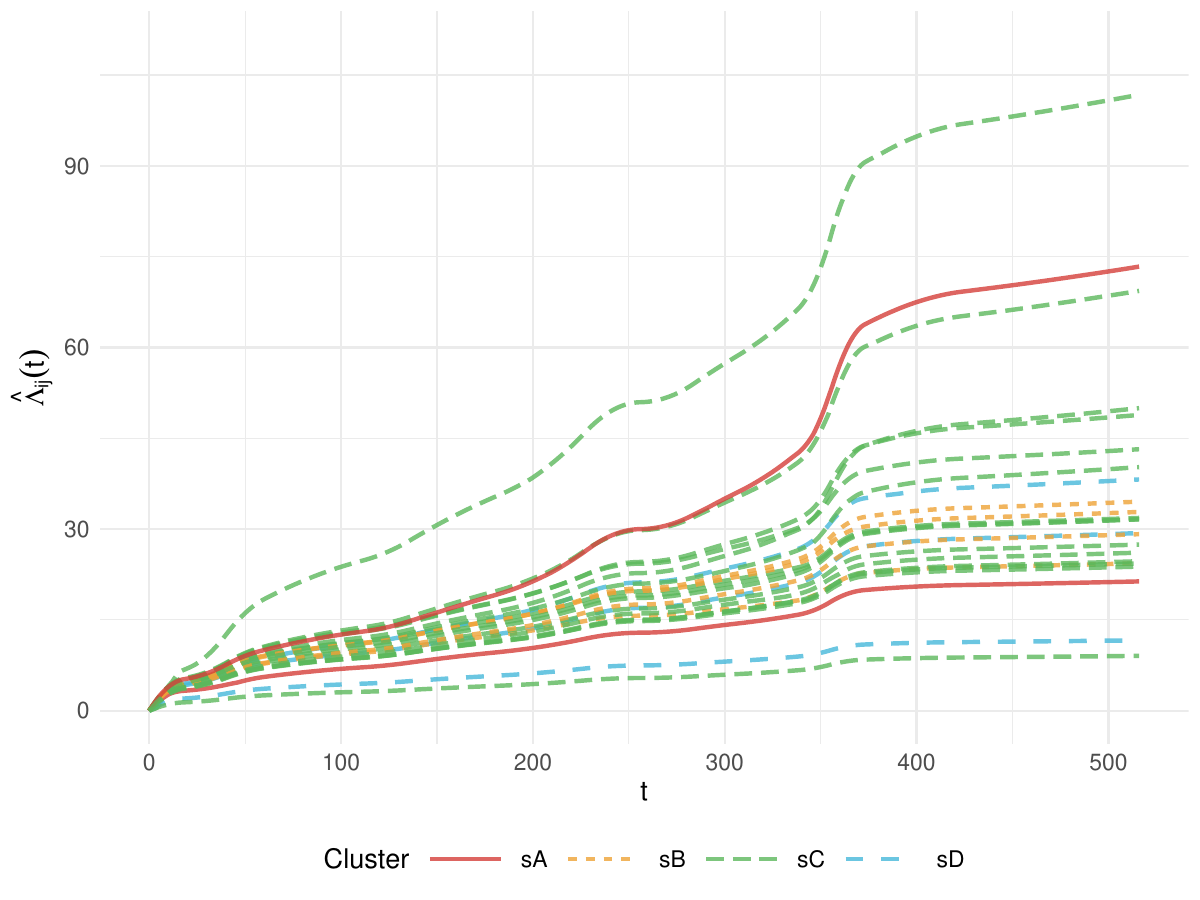}} \\
\caption{In panel (i) we show the baseline cumulative hazard of the AG model for recurrent events for the marked stochastic processes describing the dropouts. In panel (ii), we represent the reconstructed compensators as in (\ref{eq:reconstruction}) of the latter processes, each line representing a different degree \texttt{course} and each color and line type representing a different \texttt{school}, as indicated in legend.}\label{fig:casestudyCOMP}
\end{figure}

Compensators are then reconstructed as described in Eq. (\ref{eq:reconstruction}). In Figure \ref{fig:casestudyCOMP}, we display the baseline cumulative hazard and the reconstructed compensators, on two different scales. The behavior of the curves is notable: there is a steep increase in dropout counts at the beginning and end of the first year, particularly pronounced for specific degree courses. 
This pattern can be explained by several factors. Early in the first year, high dropout rates are often observed as students realize that the degree course they have chosen does not meet their expectations, leading them to switch programs or drop out. Additionally, many dropouts may occur by the end of the first year because students find the coursework too challenging or the degree program not aligned with their career aspirations. This combination of early and end-of-year dropouts contributes to the distinct peaks observed in the cumulative hazard curves.
Notable is the case of a degree course in school \texttt{sC}. 

Afterwards, the compensators are decomposed as in Eq. (\ref{eq:decompos}). The number of principal components for both levels is chosen by setting a proportion of variance explained equal to 0.99. As a result, two principal components are retained for both levels. 
At the higher hierarchical level (denoted as level 1, corresponding to the \texttt{school}), the eigenvalues obtained are $\hat{\lambda}_1^{(1)} = 74.747$ and $\hat{\lambda}_2^{(1)} = 0.434$, while at the lower hierarchical level (level 2, corresponding to the \texttt{course}), the eigenvalues are $\hat{\lambda}_1^{(2)} = 67.236$ and $\hat{\lambda}_2^{(2)} = 1.130$.

\begin{figure}
\centering
\includegraphics[width = 0.45\columnwidth]{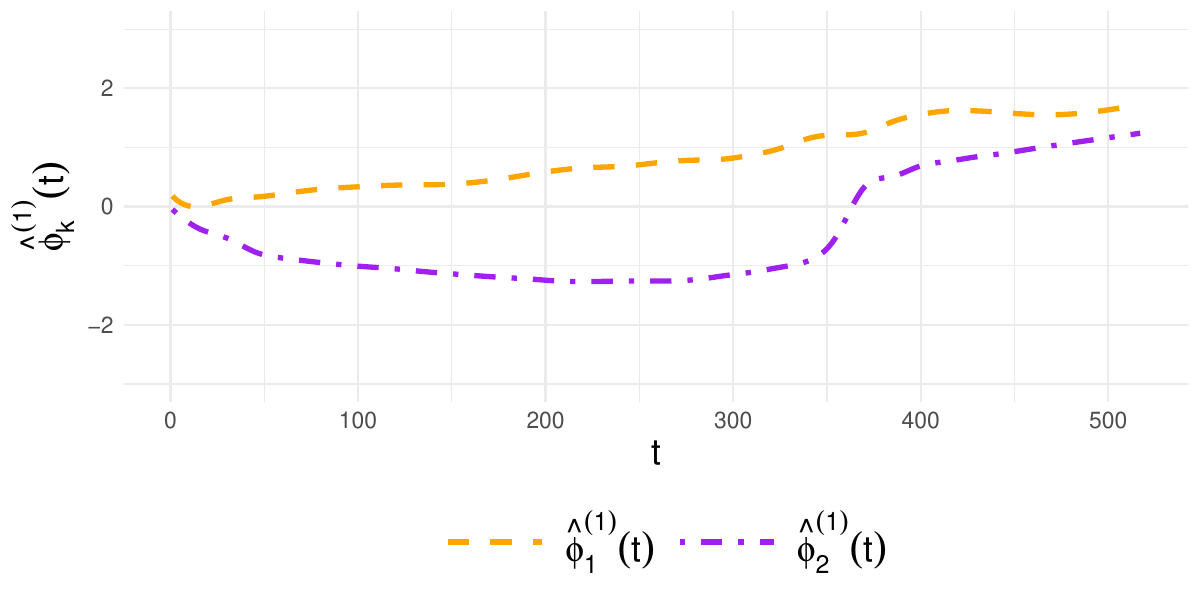} \hspace{1cm}
\includegraphics[width = 0.45\columnwidth]{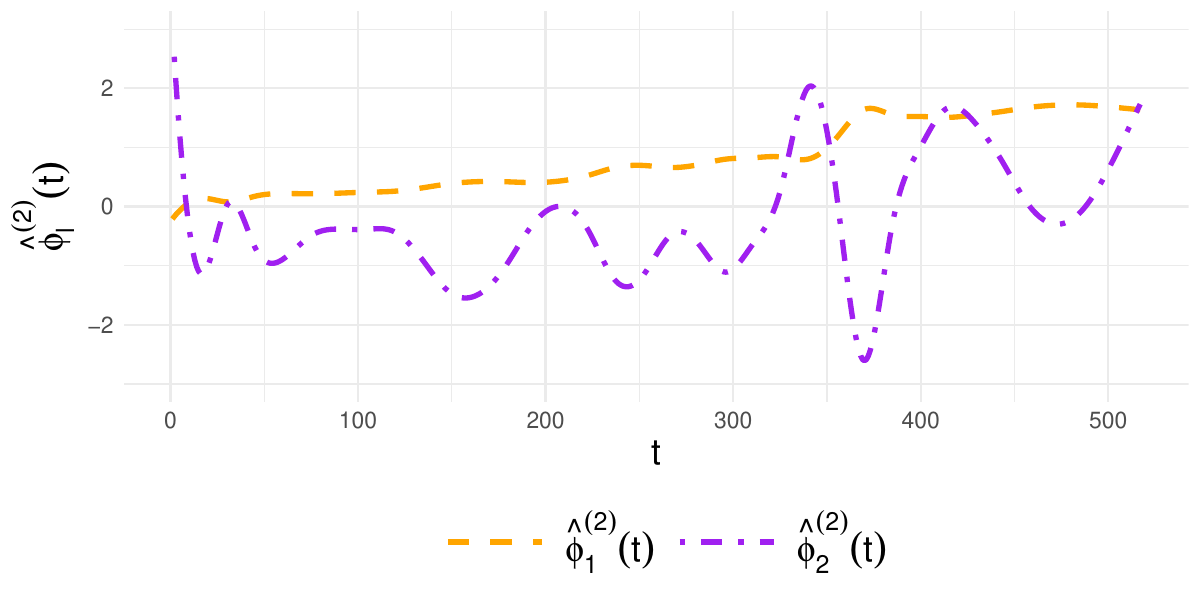} 
\caption{First two eigenfunctions at levels \texttt{school} and \texttt{course} levels (left and right panels, respectively), computed from the obtained $\hat{\Lambda}_{ij}(t)$.}
\label{fig:casestudyLambdaEigenFun}
\end{figure}

\begin{figure}
\centering
\subfloat[$1^{st}$ PC level 1 perturbation of $\hat{\mu}(t)$]{\includegraphics[width = 0.49\columnwidth]{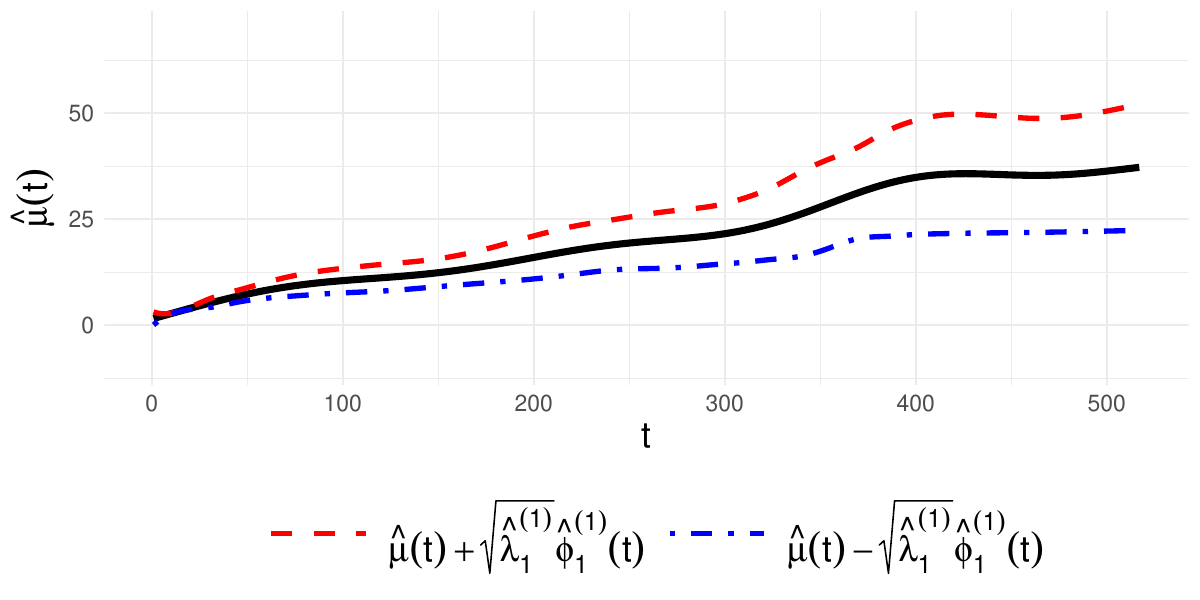}} \hspace{0.1cm}
\subfloat[$1^{st}$ PC level 2 perturbation of $\hat{\mu}(t)$]{\includegraphics[width = 0.49\columnwidth]{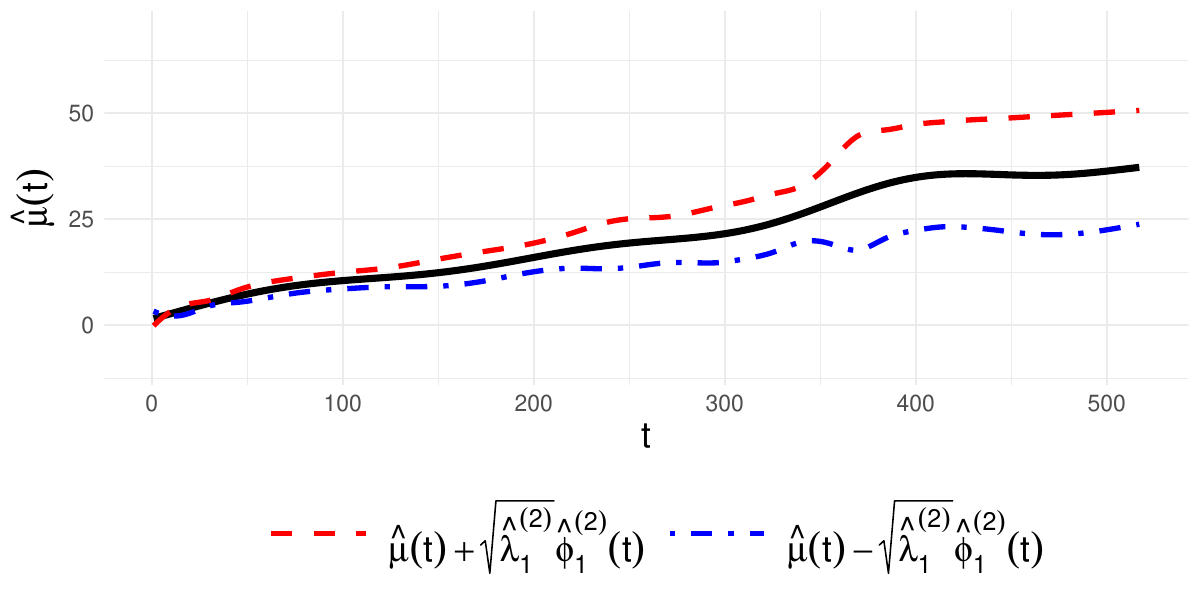}}\\
\subfloat[$2^{nd}$ PC level 1 perturbation of $\hat{\mu}(t)$]{\includegraphics[width = 0.49\columnwidth]{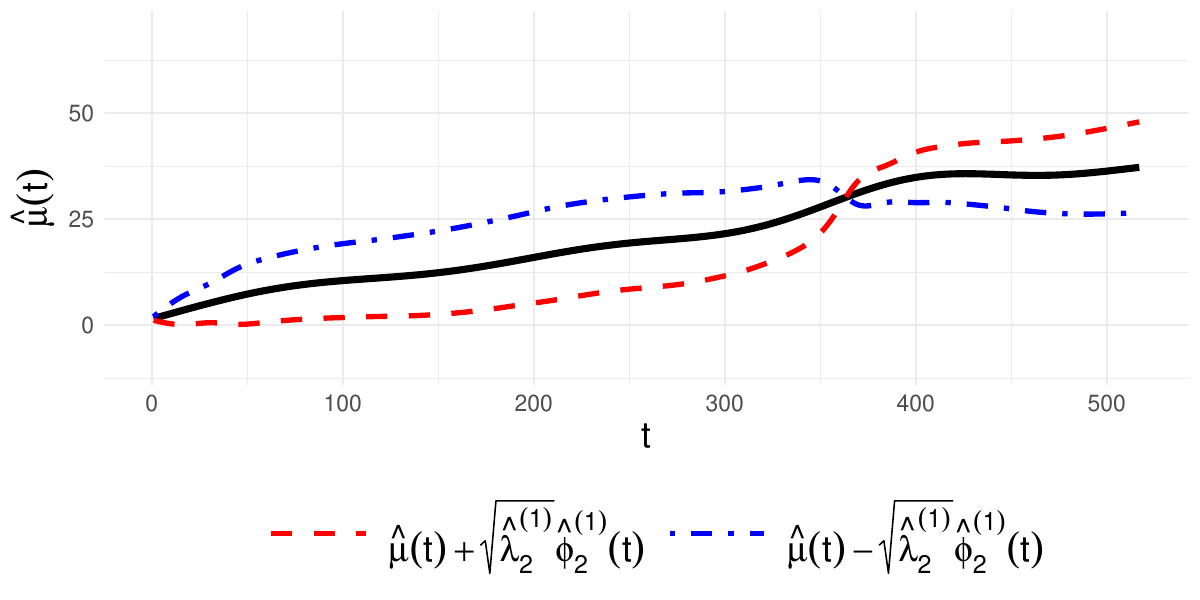}} \hspace{0.1cm}
\subfloat[$2^{nd}$ PC level 2 perturbation of $\hat{\mu}(t)$]{\includegraphics[width = 0.49\columnwidth]{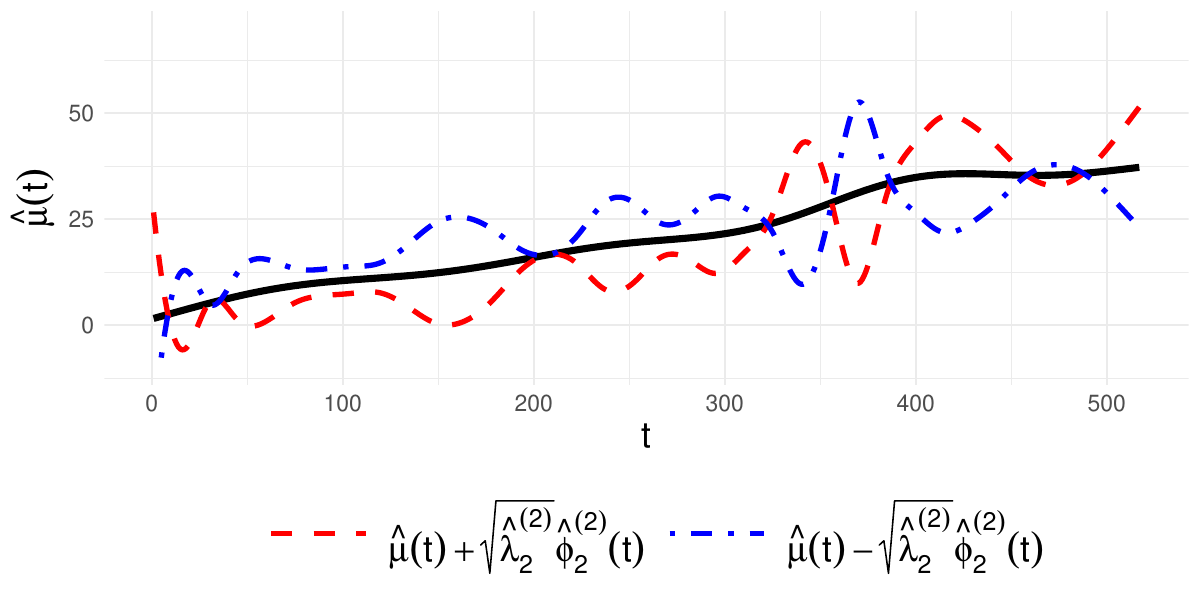}} \\
\caption{Average compensators curves $\hat{\mu}(t) = \frac{1}{n} \sum_{i,j} \hat{\Lambda}_{ij}(t)$ and their perturbations as indicated in legends, for $1^{st}$ principal component for \texttt{school} level in left panel (i) and \texttt{course} level in right panel (ii), and $2^{nd}$ principal component for \texttt{school} level in left panel (iii) and \texttt{course} level in right panel (iv).}\label{fig:casestudyPC}
\end{figure}

Figure \ref{fig:casestudyLambdaEigenFun} illustrates the first and second eigenfunctions for each of the two levels. To improve interpretability, as suggested in \cite{ramsay2005principal}, Figure \ref{fig:casestudyPC} shows the mean compensator functions $\hat{\mu}(t)$ (solid black line) along with perturbation curves (red dashed lines for positive perturbations and blue dot-dashed lines for negative) representing the eigenfunctions within one standard deviation (i.e., the square roots of the eigenvalues) from the mean, based on the MFPCA performed on $\hat{\Lambda}_{ij}(t)$.

The distribution of dropouts over time reveals distinct patterns across \texttt{school}s and degree \texttt{course}s, each associated with varying dropout risks. Notably, our analysis, as a novel contribution to the existing literature, successfully disentangles the effects of schools from those of degree programs. The first principal components at both hierarchical levels capture deviations in dropout intensity relative to the average. Specifically, schools and degree courses with a high score on the first principal component (represented by the red dashed lines) are likely to experience a higher-than-average dropout rate, while those with a low score (blue dot-dashed lines) are likely to see fewer dropouts than average. Interestingly, the dropout patterns differ between the school and course levels: at the \texttt{school} level, there is a smoother increase in dropouts toward the end of the first year (second semester), likely due to fewer fluctuations compared to the \texttt{course} level, where more variation is observed.

The second principal components, though associated with less explained variance, highlight additional temporal contrasts. At the \texttt{school} level, institutions with a high score (red dashed curve) tend to experience fewer dropouts during the first two semesters but more dropouts in the third semester. This pattern is similarly observed at the \texttt{course} level, though with greater oscillations, suggesting that some noise may also be captured. Overall, these oscillations indicate more complex dropout dynamics at the course level, where factors influencing dropouts fluctuate more over time.

\subsection{Predictive model}
\label{sec:prediction}
At this stage of the analysis, we are able to include the information derived from compensators into a predictive model (Subsection \ref{sec:logistic_casestudy}). 


Before doing so, we first outline some data preprocessing steps necessary for preparing the dataset.
We start by filtering the data to focus on \texttt{career\_start\_ay} = \lq 2017'.
To ensure consistency across observations, aligning with previous studies \cite{masci2024modelling, ragni2024assessing}, 
we create the dichotomic variable \texttt{age19} 
to denote whether students were above the age of 19 (1 if above 19, 0 otherwise).
We compute the cumulative number of CFUs 
obtained by each student by the end of the first semester, 
in the variable \texttt{ECTS1sem}.
The outcome variable \texttt{dropout3y} equals 1 if after first semester a student drops within three years, 0 otherwise.
Following preprocessing, our dataset consists of 5666 students, of which 872 dropped out.
Descriptive statistics for the over-described covariates after data pre-processing, are reported in Table \ref{tab:descr_stat_comp}, according to the dropout status. It is interesting to notice that, as expected, all numerical variables are higher when no dropout happens.
           
\begin{table}
\centering
\small
\begin{tabular}{@{}lllll@{}}
\toprule
\multicolumn{3}{@{}l}{\textbf{Variable}} & \texttt{dropout3y}=0 & \texttt{dropout3y}=1 \\\cmidrule{1-3}\cmidrule{4-5}
\textbf{Type}  & \textbf{Name} &  & {\textbf{Mean (sd)}} & {\textbf{Mean (sd)}}  \\
\midrule
\multirow{2}{*}{Numerical}  
& \texttt{admission\_score} &  & 67.00 (11.46) &  63.61 (11.62) \\
& \texttt{ECTS1sem} & & 49.79 (14.74) & 10.38 (15.39) \\
\midrule
& & \textbf{Category} &  \textbf{N (Frequency)} & \textbf{N (Frequency)} \\
\cmidrule{2-3}\cmidrule{4-5}
\multirow{15}{*}{Categorical} & \multirow{3}{*}{\texttt{origins}}   & 
       \texttt{OnSite\textsuperscript{*}} & 1033 (21.55\%)  & 232 (26.61\%)  \\
 &   & \texttt{Commuter}  & 3427 (71.49\%) &  583 (66.86\%)  \\
 &   & \texttt{Offsite}  & 334 (6.96\%)  & 57 (6.53\%)   \\
 \cmidrule{2-5}
 & \multirow{2}{*}{\texttt{gender}}  & \texttt{Male\textsuperscript{*}}   & 3200 (66.75\%)  & 670 (76.83\%)\\
 &   & \texttt{Female}  & 31594 (33.25\%) & 202 (23.17\%)  \\
  \cmidrule{2-5}
 & \multirow{4}{*}{\texttt{highschool\_type}}  & \texttt{Scientific\textsuperscript{*}}  & 3181 (66.36\%) & 542  (62.15\%) \\
 &  & \texttt{Classical}  & 327 (6.82\%) &  55 (6.31\%)    \\
 &  & \texttt{Others}    &607 (12.66\%) &  90 (10.32\%)   \\
&  & \texttt{Technical}    & 679 (14.16\%) &  185 (21.22\%)  \\
   \cmidrule{2-5}
 & \multirow{4}{*}{\texttt{income}}  & \texttt{Medium\textsuperscript{*}}  & 988 (20.61\%) & 125 (14.33\%) \\
 &  & \texttt{Grant}  & 1404 (29.29\%) &  287 (32.91\%)    \\
 &  & \texttt{High}    & 1771 (36.94\%) & 378  (43.35\%)  \\
 &  & \texttt{Low}    & 631 (13.16\%) & 82  (9.41\%)   \\
    \cmidrule{2-5}
 & \multirow{2}{*}{\texttt{age19}}  & \texttt{0\textsuperscript{*}}  & 4276 (89.19\%) &671  (76.95\%) \\
 &  & \texttt{1}  & 518 (10.81\%) & 201  (23.05\%)    \\
\bottomrule
\end{tabular}
\begin{tablenotes}
    \item\textsuperscript{*} Reference category.
\end{tablenotes}
\caption{Descriptive statistics for considered covariates after data pre-processing for \texttt{career\_start\_ay}=\lq 2017', according to the dropout status by the end of the third year.}
\label{tab:descr_stat_comp}
\end{table}

%


\subsubsection{Logistic regression model with functional compensators}
\label{sec:logistic_casestudy}

We aim to model the probability of student dropout within 3 years after the first semester (\texttt{dropout3y}), using covariates at the \texttt{studentID}-level and functional principal component scores derived from the cumulative hazard of historical dropouts over time.

We consider two principal components ($K = 2$) at the \texttt{school} level, and one principal component ($L =1 $) at the \texttt{course} level, as there is low explained variability and high oscillations in the second component that could negatively affect model performance due to noise amplification.
The binary outcome variable $Y_{ijh}$ indicating whether a student $h$ within course $j$ and school $i$ drops out within 3 years is modeled as $Y_{ijh} \sim \text{Bernoulli}(p_{ijh})$,  linear predictor given by
$$ \text{logit}(p_{ijh}) = \boldsymbol{\gamma}^T \mathbf{w}_{ijh} + \sum_{k=1}^{K} \xi_{ik} \alpha_k^{(1)} + \sum_{l=1}^{L} \zeta_{ijl} \alpha_l^{(2)} $$
for $i=1,...,I$, $j=1,...,J_i$ and $h=1,...H_{ij}$.
The vector of covariates $\mathbf{w}_{ijh}$ at the \texttt{studentID}-level includes demographic and academic information that could influence dropout risk, i.e., \texttt{origins}, \texttt{gender}, \texttt{highschool\_type}, \texttt{income}, \texttt{age19}, \texttt{admission\_score}, \texttt{ECTS1sem}.
The choice of these variables is guided by previous literature, see for instance \cite{masci2024modelling}.
Obtained results for the estimated coefficients are reported in Table \ref{tab:outputpredictivemodel}.

\begin{table}[!htbp] \centering 
\begin{tabular}{@{\extracolsep{5pt}}lccc} 
\toprule
Parameter & Estimate & Std. Error & p-value \\ 
\midrule
$\hat{\gamma}_0$ (Intercept) & 0.589 & 0.029 &  0.000 \\ 
$\hat{\gamma}_1$ (\texttt{origins} - Commuter) & 0.022 & 0.008 & 0.008 \\ 
$\hat{\gamma}_2$ (\texttt{origins} - Offsite) & -0.018 & 0.015 & 0.244 \\ 
$\hat{\gamma}_3$ (\texttt{gender} – Female) & 0.025 & 0.008 & 0.002 \\ 
$\hat{\gamma}_4$ (\texttt{highschool\_type} – Classical) & -0.000 &  0.014 & 0.988 \\ 
$\hat{\gamma}_5$ (\texttt{highschool\_type} – Others) &  0.008 & 0.012 & 0.467 \\ 
$\hat{\gamma}_6$ (\texttt{highschool\_type} – Technical) & -0.002 & 0.010 & 0.812 \\ 
$\hat{\gamma}_7$ (\texttt{income} – Grant) & -0.014 & 0.010 &  0.172 \\ 
$\hat{\gamma}_8$ (\texttt{income} – High) & 0.004 & 0.009 & 0.674 \\ 
$\hat{\gamma}_9$ (\texttt{income} – Low) &  -0.025 & 0.012 & 0.046 \\ 
$\hat{\gamma}_{10}$ (\texttt{age19} – 1) & 0.007 &  0.011 & 0.498 \\ 
$\hat{\gamma}_{11}$ (\texttt{admission\_score}) & 0.001 & 0.000  &  0.000  \\ 
$\hat{\gamma}_{12}$ (\texttt{ECTS1sem}) & -0.012  &  0.000 &  0.000 \\ 
$\hat{\alpha}_1^{(1)}$ ($\xi_{i1}$ - \texttt{school}-level score 1) & 0.002 & 0.029 &  0.049 \\ 
$\hat{\alpha}_2^{(1)}$ ($\xi_{i2}$ - \texttt{school}-level score 2) & -0.070 & 0.001  &  0.000 \\ 
$\hat{\alpha}_1^{(2)}$ ($\zeta_{ij1}$ - \texttt{course}-level score 1) & 0.001 & 0.020 & 0.083 \\ 
\bottomrule
\end{tabular} 
\caption{Estimates, standard errors, and p-values for the logistic regression model with functional compensators.} 
\label{tab:outputpredictivemodel} 
\end{table}

It is interesting to notice that coefficients related to the obtained scores at \texttt{school}-level are significant and, specifically, the one related to the first principal component is positive, indicating that the probability of dropping out within the 3 years is increased if the school in which a student is enrolled has an high score on the first principal component, and this result is coherent with the plot in Figure \ref{fig:casestudyPC} (i). 
On the other hand, the coefficient related to the the second principal component at the \texttt{school}-level is negative and statistically significant, indicating that students in schools with higher dropouts with respect to the average in the first year and lower than average in the third semester have higher probability to dropout.
At the \texttt{course}-level, the first principal component is again positively associated with dropout risk, meaning students enrolled in courses where the dropout trend is above the average are linked to an increased probability of students dropping out. 

Regarding the other covariates, \texttt{ECTS1sem} (credits earned in the first semester) is highly significant and negatively associated with dropout probability. This implies that students who pass more credits in their first semester have a lower risk of dropping out within three years. This result is in line with previous research, such as \cite{masci2024modelling} which highlights the strong predictive power of first-semester credits over later academic performance in determining dropout risk. The influence of first-semester credits is particularly pronounced, as passing more credits early on seems to provide a greater protective effect against dropout.

For the remaining covariates, the results are largely consistent with expectations, although many are not statistically significant. For example, the coefficients related to \texttt{income} and \texttt{highschool\_type} align with prior studies, but they lack statistical significance in this specific model. 
One exception is the variable \texttt{admission\_score}, which is statistically significant but reveals an unexpected positive association with dropout probability. 
Typically, one would expect that a higher admission score is related with a reduced risk of dropout, as it often indicates greater preparedness for higher education.
However, at PoliMi, students are permitted to take the admission test as early as their fourth year of secondary education. At this stage, they may not have fully developed the necessary competencies or maturity required for success in a university setting. Consequently, students who achieve high admission scores at this early stage may still struggle academically once enrolled, potentially increasing their likelihood of dropping out. This surprising result is the focus of ongoing study at PoliMi, as we aim to better understand the underlying factors contributing to this unexpected association.

In terms of model performance, the model achieves an AIC \citep{akaike1998information, bozdogan1987model} of 808.68, and excellent predictive power, with an AUC \citep{hanley1982meaning} of 0.9425 and an accuracy of 0.92. Sensitivity (0.954) and precision (0.951) are both high, indicating that the model is very good at correctly identifying students who are at risk of dropping out. Specificity (0.732), though slightly lower, still indicates a reasonable ability to identify students who are not at risk. These performance metrics suggest that the model is well-calibrated for predicting dropout risk, with a particular strength in identifying students at higher risk.

If we fit the same model but exclude the compensators' information, we obtain an AIC of 823.85 and an AUC of 0.9418. This indicates that the model provides a better fit when compensator information is included.
On the other hand, while a mixed-effects model can account for unobserved heterogeneity, it primarily introduces scaling factors and may not be as suitable in our case. It lacks the capacity to capture the intricate temporal dynamics that our compensator-based approach, combined with multilevel functional principal component analysis, effectively models over time. Also, the authors in \cite{baraldo2013outcome} compared such models, demonstrating that mixed-effects models do not offer superior performance.



\section{Discussion}
\label{sec:discussion}

Addressing student dropouts is a critical concern for universities, both academically and financially. Each dropout represents an inefficient use of institutional resources allocated to recruitment, teaching, and student support. Reducing dropout rates directly impacts both financial stability and the overall effectiveness of educational systems.

One of the complexities in tackling this issue lies in the heterogeneous nature of dropout behaviour across degree programs and schools. Different academic disciplines present unique challenges - some programs may experience high dropout rates early on due to demanding foundational courses, while others see increased dropouts as students' careers progress. Similarly, the dropout patterns can vary considerably across schools within the same university, influenced by factors such as faculty engagement and available student support.

In this paper, we present a novel approach to modelling dropout behaviour by examining occurrences over time within both degree programs and schools. 
Our work has two main goals: (i) to estimate the dropout trends over time and examine its variability across different degree programs and schools, and (ii) to leverage this information in a predictive framework at the student level.
To achieve these objectives, we utilize Cox-based regression for recurrent events to capture the temporal dynamics and underlying structure of dropout trends. In this initial phase, we employ an AG model, as supported by existing literature \citep{spreafico2021functional}.
However, it is important to note that other modeling choices, such as those proposed by \cite{baraldo2013outcome}, which build on \cite{pena2007semiparametric}, are also possible. Selecting the appropriate model can be challenging; consequently, this first step of the analysis could be replicated using alternative modeling approaches, allowing for further exploration and validation of our findings.
By decomposing dropout patterns within programs and schools through multilevel functional principal component analysis, we provide a detailed view of critical time periods when dropout rates tend to spike. This approach offers both visual and quantitative insights into when students are most at risk of leaving their studies, allowing institutions to identify vulnerable cohorts and periods. Furthermore, by capturing these temporal trends, we gain a deeper understanding of how dropout behaviour varies across disciplines.

Our predictive model adds significant value by incorporating historical dropout data on current dropout behaviour. By integrating information from previous cohorts, our approach allows universities to more accurately forecast future dropout risks and target proactive interventions. This enables educational institutions to identify at-risk students earlier in their academic journeys, based on a combination of baseline characteristics such as academic performance in first semester, socioeconomic status, and previous schools attended. With this information, universities can implement more personalized support strategies aimed at reducing dropouts, such as tutoring classes, thus improving student retention and overall success.

While our model presents promising results, there are several limitations and areas for further development. First, this is a preliminary analysis, and the results should be validated across multiple academic years to ensure robustness. Cross-validation techniques could be employed to improve the stability and generalizability of the model outcomes. Additionally, the impact of external factors like the Covid-19 pandemic - which may have fundamentally altered student engagement and retention - should be incorporated into future analyses. Understanding how the pandemic influenced dropout patterns could further refine our predictions.
Furthermore, our analysis focuses on the first three semesters, a period selected because highly predictive of dropout risk. Since the compensator must be integrated over historical data, extending the observation period beyond this point did not yield significant improvements in the analysis; however, this remains an area for further evaluation in future research.
Indeed, while considering this period in the compensators reconstruction allows capturing early dropouts at course and school levels, future extensions could consider the entire three-year duration of undergraduate programs to provide a more comprehensive understanding of student retention. Incorporating time-to-event data would allow us to model dropout risk more accurately over time, addressing both whether and when students are likely to drop out.

\section*{Competing interests}
No competing interest is declared.

\section*{Data availability}
The participants of this study did not give written consent for their data to be shared publicly, so due to the sensitive nature of the research, the full supporting data is not available.

\section*{Acknowledgements}
The authors acknowledge the support by MUR, Italy,
grant \lq Dipartimento di Eccellenza 2023-2027\rq.

\begin{appendix}
\counterwithin{equation}{section}
\counterwithin{figure}{section}
\counterwithin{table}{section}
\renewcommand\theequation{\thesection\arabic{equation}}

\section{Compensators reconstruction}
\label{app:compreconstr}

The realizations of each compensator $\Lambda_{ij}(t)$ for each unit $j$ in cluster $i$, employing result in Eq. (\ref{Eq:AGmodels}) can be expressed as follows:
\begin{align*}
    \Lambda_{ij}(t) & = \int_0^t  Y_{ij}(s) \lambda_0(s) \; \text{exp}(\boldsymbol{\beta}^T \mathbf{x}_{ij}(s) + \boldsymbol{\theta}^T \mathbf{z}_{ij}(s)) ds  \nonumber \\
    & = \sum_{k=0}^{N_{ij}(t^-)}  \int_{t_k^{(ij)}}^{t_{k+1}^{(ij)} \wedge t} \lambda_0(s) \; \exp(\boldsymbol{\beta}^T \mathbf{x}_{ij}(s) + \boldsymbol{\theta}^T \mathbf{z}_{ij}(s)) ds  \nonumber \\
    & \simeq \sum_{k=0}^{N_{ij}(t^-)} \; \exp(\boldsymbol{\beta}^T \mathbf{x}_{ij}(t_k^{(ij)}) + \boldsymbol{\theta}^T \mathbf{z}_{ij}(t_k^{(ij)})) \int_{t_k^{(ij)}}^{t_{k+1}^{(ij)} \wedge t} \lambda_0(s)  ds  \nonumber \\
    & =  \sum_{k=0}^{N_{ij}(t^-)} \; \exp(\boldsymbol{\beta}^T \mathbf{x}_{ij}(t_k^{(ij)}) + \boldsymbol{\theta}^T \mathbf{z}_{ij}(t_k^{(ij)})) 
    \big[ \tilde{\Lambda}_0(t_{k+1}^{(ij)} \wedge t) - \tilde{\Lambda}_0(t_k^{(ij)})\big] 
\end{align*}
where $a\wedge b = \min\{a,b\}$, $N_{ij}(t^-)$ represents the number of occurrences that have happened strictly before time $t$, and $\hat{\boldsymbol{\beta}}$ and $\hat{\boldsymbol{\theta}}$ are the estimated vectors of coefficients.

\end{appendix}

\bibliographystyle{chicago} 
\bibliography{bibliography}

\end{document}